\documentclass[12pt,twocolumn]{article}

\usepackage[colorlinks,urlcolor=blue,linkcolor=blue,citecolor=blue]{hyperref}
\usepackage{graphicx}
\usepackage{amsmath, amssymb, amsfonts}
\usepackage{booktabs}
\usepackage{longtable}
\usepackage{adjustbox}
\usepackage{float}
\usepackage{subfig}
\usepackage{array}
\usepackage{makecell}
\usepackage[dvipsnames]{xcolor}
\usepackage{tikz}
\usetikzlibrary{fit,shapes.geometric,positioning,backgrounds,calc,arrows.meta,mindmap,shadows}
\usepackage{titlesec} 

\usepackage{authblk} 
\titleformat{\section}{\normalfont\large\bfseries}{\thesection}{1em}{}

\newcommand\blfootnote[1]{%
    \begingroup
    \renewcommand\thefootnote{}%
    \setlength{\parindent}{0pt}
    \footnote{#1}%
    \addtocounter{footnote}{-1}%
    \endgroup
}

\title{\LARGE The Rise of AI-Generated Anime Avatars: \\ Trends, Challenges, and Opportunities}

\author[1]{Fernanda Miyuki Yamada}
\author[2]{João Paulo Gois}
\author[1]{Hiroki Takahashi}

\affil[1]{The University of Electro-Communications, Chofu, Tokyo, Japan}
\affil[2]{Federal University of ABC, Santo André, São Paulo, Brazil}

\date{}

\begin{document}

\twocolumn[
\vspace*{-6em}
\maketitle
\begin{abstract}
{\small The rise of 3D anime-style avatars in gaming, virtual reality, and other digital media has driven significant interest in automated generation methods capable of capturing their distinctive visual characteristics. These include stylized proportions, expressive features, and non-photorealistic rendering. This paper reviews the advancements and challenges in using deep learning in 3D anime-style avatar generation. We analyze the strengths and limitations of these methods in capturing the aesthetics of anime characters and supporting customization and animation. Additionally, we identify and discuss open problems in the field, such as difficulties in resolution and detail preservation, and constraints regarding the animation of hair and loose clothing. This article aims to provide a comprehensive overview of the current state-of-the-art and identify promising research directions for advancing 3D anime-style avatar generation.}
\end{abstract}
\vspace*{1em}
]

\blfootnote{© 2026 IEEE. Personal use of this material is permitted. Permission from IEEE must be obtained for all other uses, in any current or future media, including reprinting/republishing this material for advertising or promotional purposes, creating new collective works, for resale or redistribution to servers or lists, or reuse of any copyrighted component of this work in other works. \\ DOI: 10.1109/MCG.2025.3627323}
\section{Introduction}

\begin{figure}[ht!]
    \centering
     \subfloat[\centering Happy anime girl]{{\includegraphics[width=0.2\textwidth]{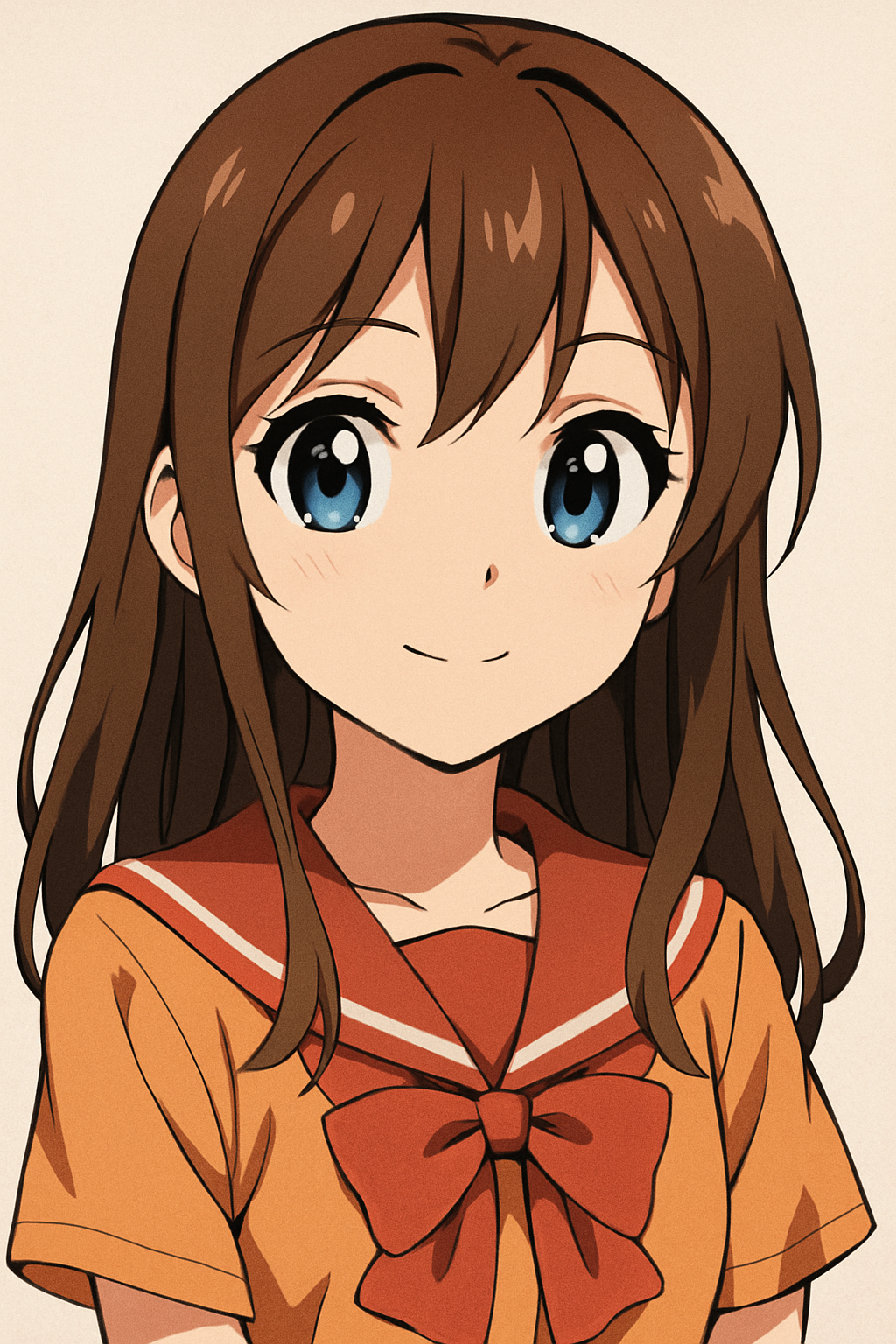} }}%
    \subfloat[\centering Upset anime boy]{{\includegraphics[width=0.2\textwidth]{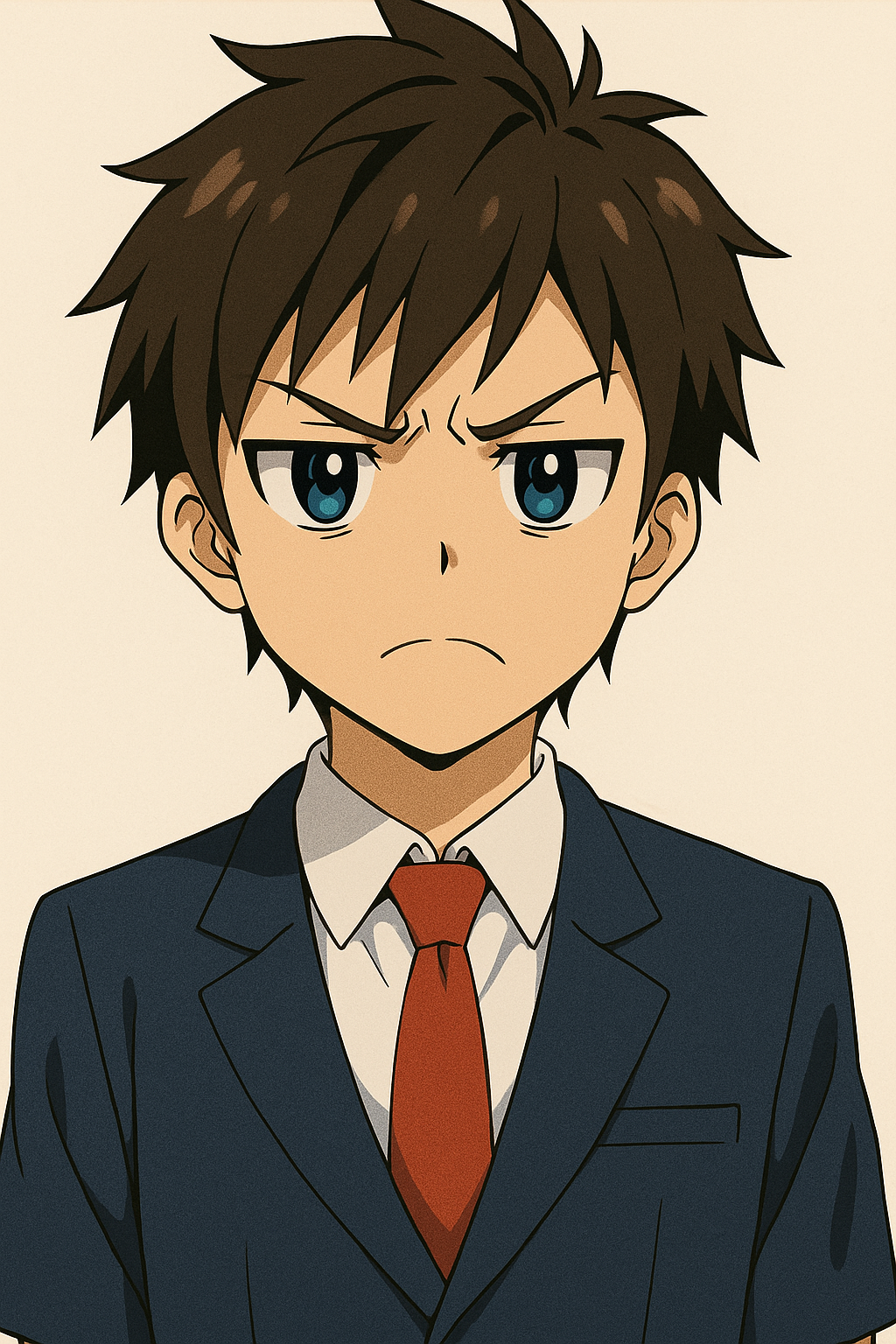} }}%
    \caption{Anime art style generated using GPT-4o by OpenAI. Characters are typically characterized by large, expressive eyes, small noses, and simplified facial features that convey a wide range of emotions. }
    \label{anime}
\end{figure}

Anime represents a compelling and vibrant sector of animated media originated in Japan, well-known for its distinct artistic conventions such as large, expressive eyes, simplified facial features, and exaggerated proportions. Anime is a reflection of the Japanese culture, having strong ties to politics, history, economy, gender, religion, and social
beliefs. Its influence is not limited to Japan either, with a growing trend of global collaboration, as international studios increasingly co-produce shows alongside Japanese creators. Numerous studies have linked anime to cross-cultural practices, highlighting its role in boosting tourism to Japan and sparking increased interest in learning the Japanese language. Figure~\ref{anime} presents characters of the anime art style, showcasing the stylized proportions associated with the genre.

\begin{figure*}[ht!]%
    \centering
     \subfloat[\centering Generative framework for 3D anime-style avatars]{{\includegraphics[width=0.52\textwidth]{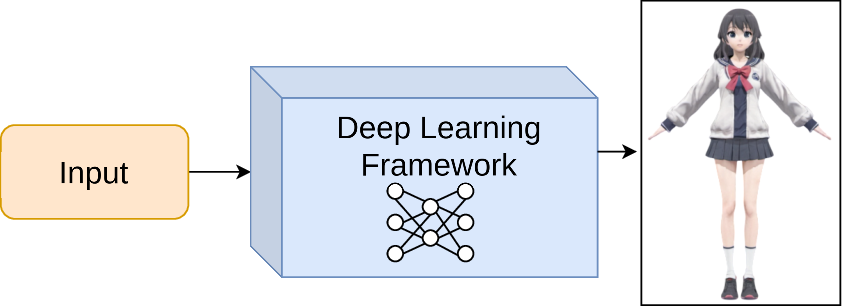} }}%
    \vspace{0.1in}
    
    \subfloat[\centering Text prompt]{{\includegraphics[width=0.27\textwidth]{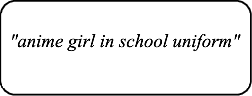} }}%
    \quad
    \subfloat[\centering Single-view image]{{\includegraphics[width=0.18\textwidth]{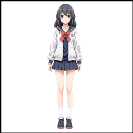} }}%
    \quad
    \subfloat[\centering Multi-view image]{{\includegraphics[width=0.33\textwidth]{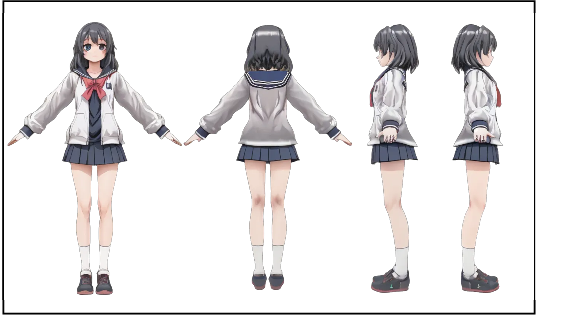} }}%

   \vspace{0.1in}
    \caption{ General pipeline for 3D anime-style avatar generation. The framework (a) is designed for a specific input type: (b) text prompt, (c) single-view image, or (d) multi-view image. }
    \label{flow}
\end{figure*}

The global expansion of anime has also increased the demand for 3D anime-style avatars, especially with the rise of Metaverse applications and the Virtual YouTuber (VTuber) industry. Recent advances in AI-driven approaches, particularly deep learning, enable the generation of 3D avatars from accessible inputs, such as images and text. Despite these advances, a gap remains in the field. Most methods focus on 3D realistic human avatar generation, neglecting anime-style representations. Approaches dedicated to 3D realistic human avatars often use text prompts as inputs and produce high-quality, detailed outputs. In contrast, anime-style counterparts usually rely on simpler networks and use image-based inputs. This reliance on image data creates a significant drawback, as it requires extensive datasets for training and depends on some level of artistic skill, further limiting accessibility and flexibility. Figure~\ref{flow} shows the input types and the 3D avatar generation process using a deep learning framework.

This article presents an overview of deep learning approaches for 3D anime-style avatars. We highlight how frameworks dedicated to 3D realistic human avatars struggle with exaggerated proportions and aesthetics, and explore opportunities for developing frameworks dedicated to anime-style. In addition to summarizing current techniques, we identify key limitations in data availability, model design, and evaluation protocols. Our goal is to provide a clear roadmap of the current landscape and outline future directions for researchers and developers interested in the field.

\section{3D Realistic Human Avatar Generation}

We begin by analyzing recent advances in 3D realistic human avatar generation, as this area has been more thoroughly explored in recent years, demonstrating remarkable progress in diffusion-based generation. Notably, recent methods in this field have converged on text-prompt-based approaches, allowing users to describe the appearance of 3D avatars through natural language descriptions.

Recent advances in 3D realistic human avatar generation have concentrated on enhancing visual quality and structural consistency, with DreamAvatar~\cite{cao2024dreamavatar} pioneering a dual-observation-space design that addressed fundamental challenges such as the Janus problem, where the model produces faces with conflicting front and back features, through innovative canonical and posed space connections. Another major direction aims to create rigged avatars suitable for animation and interactive use. DreamWaltz~\cite{huang2023dreamwaltz} combined diffusion models with Neural Radiance Fields (NeRF) and the Skinned Multi-Person Linear Model (SMPL)~\cite{loper2023smpl} model for skeletal guidance. Later methods extended this with  SMPL eXpressive (SMPL-X)~\cite{pavlakos2019expressive} to improve facial motion and details.

Current methods represent sophisticated integrations of multiple advances. The combination of three core components for the backbone of text-to-3D avatar generation has proven particularly successful in generating 3D realistic human avatars with high-resolution: The first is diffusion models that serve as text-to-image architectures to guide the visuals of the generated 3D avatar. It interprets textual or semantic input to produce 2D visual references that define the desired look of the avatar under different viewpoints. The second is ControlNet, which provides enhanced pose control and ensures consistency across various sampled viewpoints, allowing precise adjustment of the appearance and positioning of the 3D model. It serves a role similar to pose-driven deformation in traditional animation pipelines, as it ensures that the generated content conforms to a specific skeletal structure or motion sequence, even though the image itself is generated by a neural process. Finally, parametric human models, especially SMPL and SMPL-X, ensure realistic proportions while supporting animation. 

\begin{figure*}[ht!]
    \centering
    \includegraphics[width=0.9\linewidth]{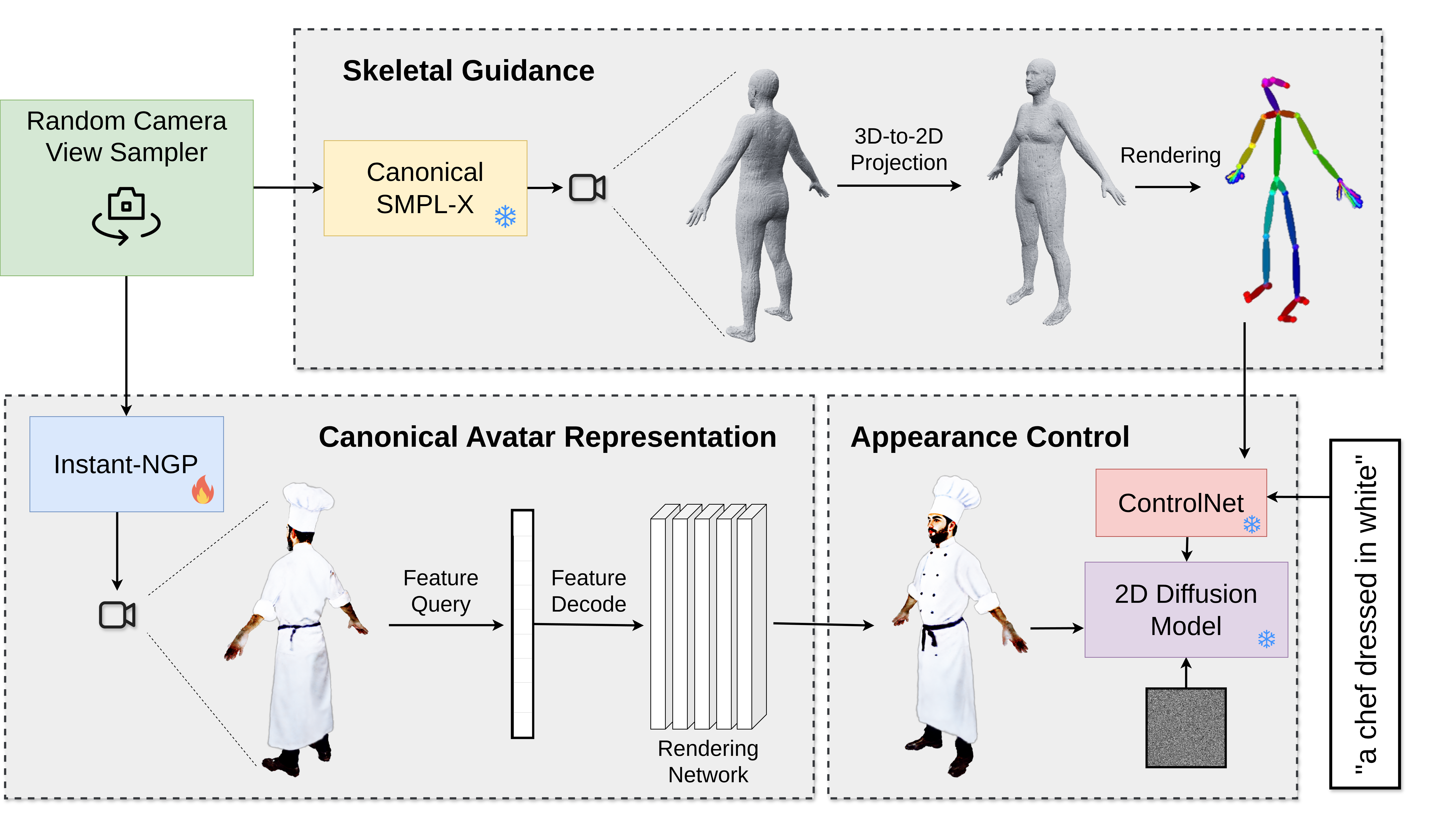}
    \caption{General pipeline for 3D realistic human avatar generation. The process starts with an SMPL-X parametric model rendered into a skeletal keypoint representation, which guides ControlNet to condition a 2D diffusion model. Instant-NGP generates a NeRF representation of the 3D avatar, whose appearance is guided by the output of the 2D diffusion model. }
    \label{controlnet}
\end{figure*}

DreamWaltz-G~\cite{huang2024dreamwaltzg} incorporates this backbone through a multi-stage generation process. While the backbone offers a foundation for skeletal guidance and appearance control during the generation process, the framework employs NeRF reconstruction via Instant Neural Graphics Primitives (Instant-NGP)~\cite{huang2024dreamwaltzg} to synthesize multiple generated images into a continuous 3D representation with view-consistent appearance. The animation rigging stage then converts the NeRF output into a 3D Gaussian Splatting representation, preserving fine-grained surface and texture details while enabling efficient rendering and interactive editing capabilities. Figure~\ref{controlnet} illustrates this general pipeline architecture, which underlies DreamWaltz-G and other frameworks such as DreamWaltz~\cite{huang2023dreamwaltz}.

Despite being developed for 3D realistic human avatar generation, these methods exhibit some level of adaptability to different artistic styles. Different frameworks have validated their approach on anime and cartoon-style characters. However, when directly applied to 3D anime-style avatar generation, text-prompt-based approaches face inherent challenges in capturing the stylistic complexity of anime aesthetics. This issue originates from the fact that most text-to-image models are trained predominantly on photographs of real humans, resulting in priors that are biased toward realistic humans.

\section{3D Anime-Style Avatar Generation}\label{survey}

The frameworks dedicated to generating 3D anime-style avatars use image data as input, either multi-view or single-view images of an anime character. These methodologies represent distinct strategies for addressing the unique challenges of 3D anime-style avatar generation, each offering different trade-offs between input requirements, generation quality, and practical applicability. Single-view methods offer fine control over 3D avatar appearance with less data but face challenges in unseen regions and self-occlusion, while multi-view methods improve geometric accuracy and detail but require more data and consistent reference images.

\begin{figure}[ht!]
    \centering
     \includegraphics[width=0.4\textwidth]{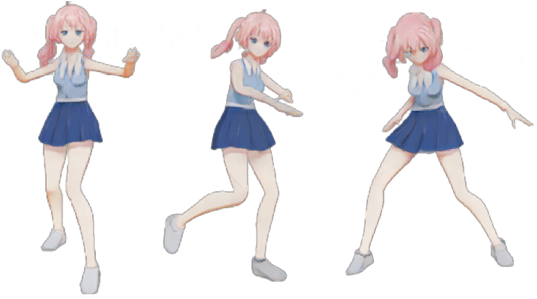} %
    \caption{3D anime-style avatar generated by CoNR~\cite{lin2022collaborative} presented in different poses, licensed under MIT License.}
    \label{conr}
\end{figure}

The first major step in this area was made by Lin et al.~\cite{lin2022collaborative}, who introduced Collaborative Neural Rendering (CoNR). Their framework bridges the gap between 2D anime-style image generation and 3D modeling through multi-view reconstruction. Their framework introduces Ultra-Dense Pose (UDP) mapping, a novel technique that establishes a correspondence between 3D landmarks and body surfaces within 2D images. This approach uses convolutional image-based networks to generate character images in specified poses, while employing feature space cross-view dense correspondence to enhance rendering quality. The main contribution lies in their systematic approach to pose standardization, addressing common challenges in pose variation and occluded views. This standardization is facilitated by a UDP Detector that estimates landmarks and masks from input images, creating a robust approach for pose transfer and animation generation. The authors provide a dataset containing over 700,000 samples of different characters, each paired with UDP annotations, a significant resource that has influenced subsequent research directions. The limitations observed in CoNR, including the reliance on multiple views and the associated data acquisition challenges, motivated a fundamental shift toward single-view reconstruction approaches. Figure~\ref{conr} shows a 3D anime-style avatar generated by CoNR~\cite{lin2022collaborative} in different poses.

\begin{figure}[ht!]
    \centering
     \includegraphics[width=1\linewidth]{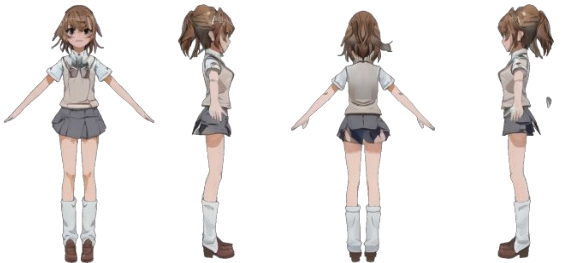}
    \caption{Different views of a single 3D anime-style avatar generated by CharacterGen~\cite{peng2024charactergen}, licensed under CC BY-SA 4.0.}
    \label{charactergen}
\end{figure}

Peng et al.~\cite{peng2024charactergen} introduced CharacterGen, which represents a significant leap in addressing data accessibility concerns while maintaining generation quality. The framework tackles the persistent challenges of diverse body poses and self-occlusion through a sophisticated two-stage process, demonstrating notable improvements in detail preservation compared to earlier methods. The first stage performs viewpoint lifting, transforming a single input image into multiple viewpoints while simultaneously converting the input pose into a canonical form. Rather than relying on parametric models, the system extracts 2D keypoints from input images, providing structural guidance for 3D avatar generation in A-pose. While this approach offers simplicity suitable for lightweight applications, it trades off the shape and pose parameterization benefits that parametric models provide for animation and real-time editing capabilities. The second stage employs a transformer-based reconstruction model to create detailed 3D model meshes. Trained on the Anime3D dataset, CharacterGen achieves remarkable efficiency, generating characters in less than one minute. Figure~\ref{charactergen} shows different views of a 3D anime-style avatar generated by CharacterGen~\cite{peng2024charactergen}.

\begin{figure}[ht!]
    \centering
     \includegraphics[width=0.7\linewidth]{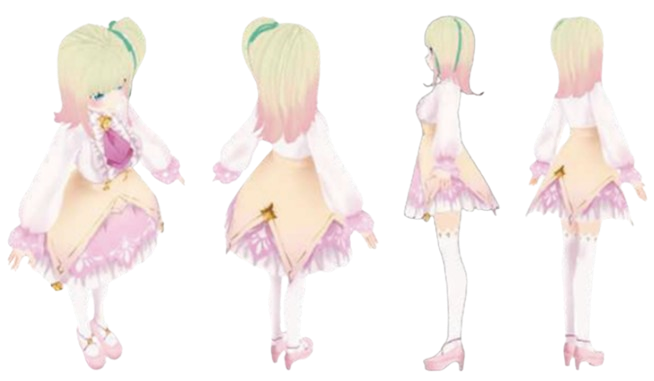}
    \caption{Dynamic perspective views of a single 3D anime-style avatar generated by NOVA-3D~\cite{wang2024nova}, licensed under CC BY-SA 4.0.}
    \label{nova}
\end{figure}

Although multi-view approaches offer accurate reconstructions, Wang et al.~\cite{wang2024nova} pointed out that their dependence on carefully aligned inputs restricts usability in real-world settings. Their response, NOVA-3D, represents a strategic refinement of the multi-view approach, addressing this challenge by developing a framework capable of reconstructing 3D anime characters from non-overlapping views. While CoNR requires four views as input, NOVA-3D dramatically reduces data requirements by utilizing two non-overlapping views, specifically back and front images that share no common regions. The framework uses the proposed NOVA-Human dataset, containing 10.2k 3D anime character models with multi-view images and camera parameters, focusing on A-poses. The authors propose their reconstruction process as a volumetric generative network conditioned on non-overlapped images. NOVA-3D uses a dual-perspective feature extraction module that analyzes the front and back images to capture the shape and appearance of the character from multiple angles. However, this efficiency comes with acknowledged limitations, with authors noting constraints in scenarios where front and back views are unavailable and recognizing that the potential for customization and rigged animation remains underexplored. Comparative analysis reveals that while NOVA-3D reduces input requirements, the generated 3D avatars tend to exhibit less detail than those produced by CoNR, particularly in facial features and clothing textures. Figure~\ref{nova} shows different views of a 3D anime-style avatar generated by NOVA-3D~\cite{wang2024nova}.

\begin{figure}[ht!]
    \centering
     \includegraphics[width=1\linewidth]{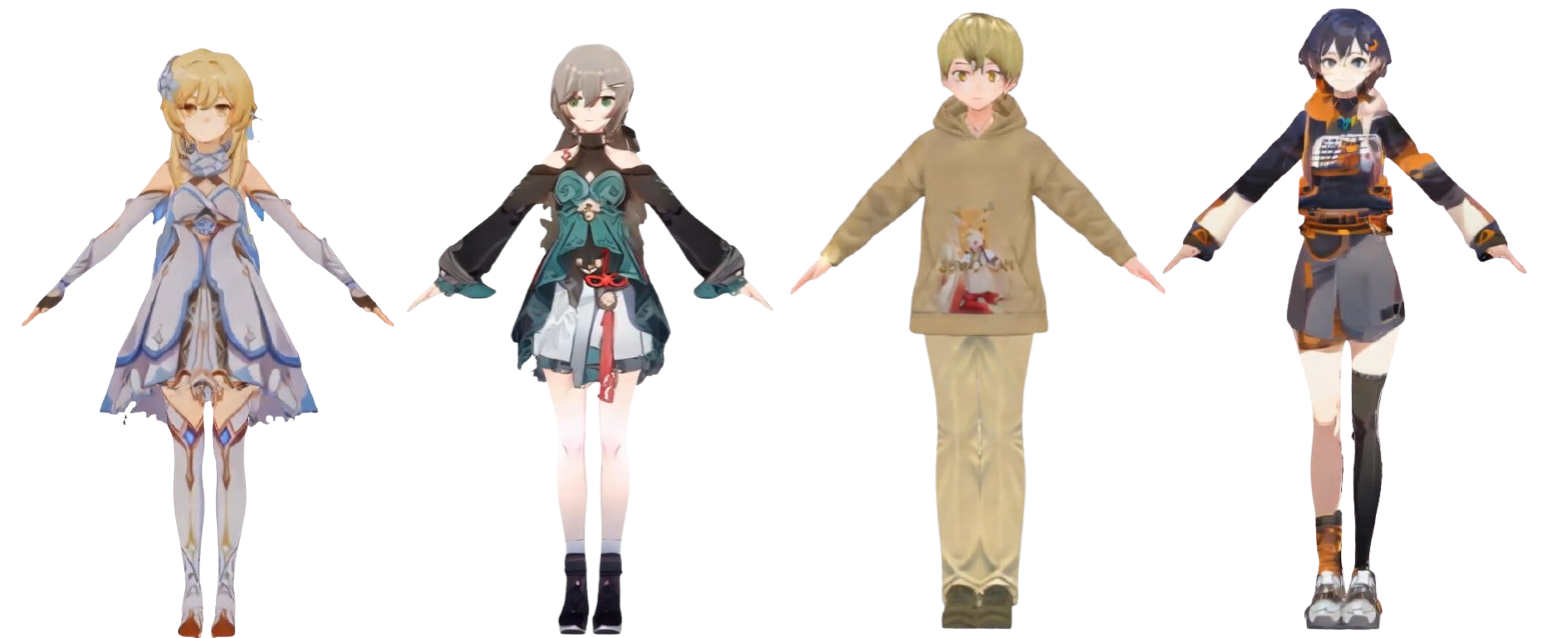}
    \caption{Frontal view of 3D anime-style avatars generated by StdGEN~\cite{he2024stdgen}, licensed under Apache License 2.0. Image has been modified.}
    \label{stdgen}
\end{figure}

The success of CharacterGen in single-view reconstruction opened new avenues for enhancing the semantic understanding of 3D avatar generation. He et al.~\cite{he2024stdgen} advanced this direction with StdGEN, introducing a shift toward semantic decomposition of 3D avatar components that builds upon the single-view approaches while addressing their limitations in component-level control. Unlike previous methods that treat 3D avatars as monolithic entities, StdGEN recognizes and processes distinct components such as hair, body, and clothing as separate attributes. This approach integrates semantic attributes directly into the reconstruction process, enabling the model to interpret high-level character descriptions and translate them into corresponding visual features. A particular contribution is the ability to apply multi-layer mesh refinement to each 3D avatar component separately, enabling unprecedented control over individual elements, such as hair and clothing. However, the authors acknowledge limitations in resolution quality and potential requirements for manual editing. Additionally, the average generation time is three times longer than CharacterGen, highlighting a clear trade-off between semantic awareness and efficiency. Figure~\ref{stdgen} shows some 3D anime-style avatars generated by StdGEN~\cite{he2024stdgen}.

These four approaches collectively illustrate the evolution from comprehensive multi-view reconstruction toward increasingly accessible single-view approaches, with each advancement addressing specific limitations while introducing new challenges. The progression from CoNR four-view requirement to NOVA-3D dual-view approach demonstrates ongoing efforts to balance quality with practicality within the multi-view approach, while the development from CharacterGen efficient single-view reconstruction to StdGEN semantic-aware approach shows the field movement toward a more sophisticated understanding of 3D avatar structure and composition. The multi-view approaches offer superior detail preservation and geometric accuracy by using multiple perspectives, but require more complex data acquisition and processing pipelines. Single-view approaches prioritize accessibility and ease of use, requiring only a single input image while employing sophisticated inference techniques to hallucinate missing information, but still face persisting challenges in reconstructing occluded or ambiguous regions from limited visual information.

\begin{figure}[ht!]%
    \centering
     \subfloat[\centering ``an anime boy'']{{\includegraphics[width=0.138\textwidth]{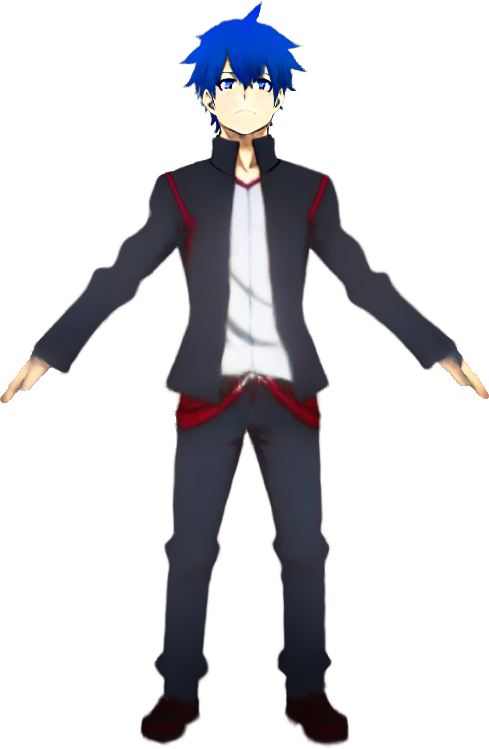} }}%
     \quad
    \subfloat[\centering ``an anime girl in a sports jersey'']{{\includegraphics[width=0.138\textwidth]{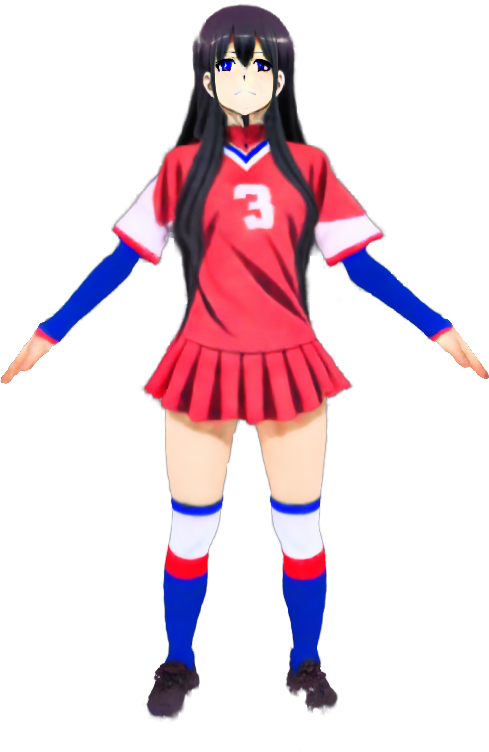} }}%
    \quad
    \subfloat[\centering ``an anime girl in a kimono'']{{\includegraphics[width=0.138\textwidth]{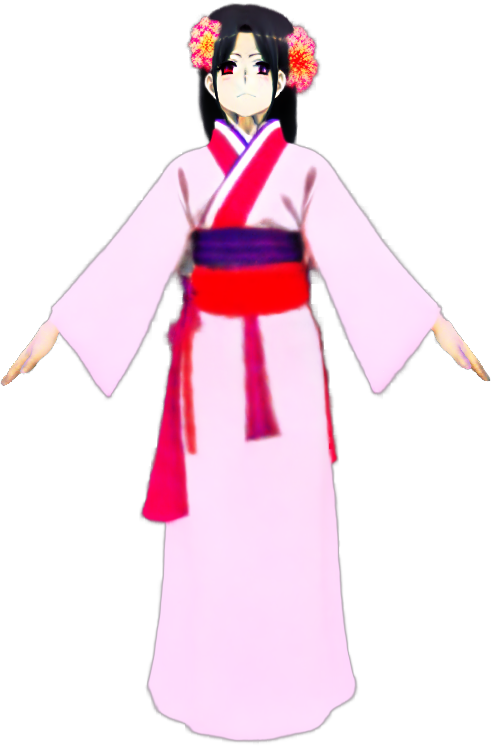} }}%

    \caption{Frontal view of 3D anime-style avatars generated from text prompts by AniDream~\cite{yamada2025anidream}, licensed under MIT License.}
    \label{anidream}
\end{figure}

Recently, Yamada \& Takahashi proposed AniDream~\cite{yamada2025anidream}, introducing a fundamentally different approach as the first framework to generate anime avatars directly from text prompts. Rather than requiring users to provide reference images, it allows creators to simply describe the appearance of the 3D anime-style avatar in natural language. Drawing inspiration from DreamWaltz-G~\cite{huang2024dreamwaltzg} and building on the pipeline illustrated in Fig.~\ref{controlnet}, AniDream adds two key improvements specifically for anime style. AniDream introduces a fine-tuned adaptation module that can efficiently adjust the diffusion model to capture anime-specific features and a loss function that encourages the approach to produce a cel-shading effect. Experimental results show that AniDream produces 3D avatars that look distinctly anime-styled while maintaining accurate body proportions. Figure~\ref{anidream} presents avatars created using only text prompts~\cite{yamada2025anidream}.

The literature shows that there are fundamental trade-offs in 3D avatar generation: input complexity versus output quality, data requirements versus accessibility, and geometric accuracy versus inference sophistication. A common challenge among image-based approaches is that the generated 3D models often suffer from detail loss, visible as overly smoothed surfaces, missing sharp contours, or distorted local structures, leading to low-resolution approximations of the character represented in the input data. Text-prompt-based approaches can potentially address this limitation by exploiting models trained on millions of images, allowing them to draw upon broader visual knowledge rather than being constrained by the details visible in a reference image.

\section{Trends and Challenges} \label{discussion}

As previously stated, there is a noticeable trend in focusing on 3D realistic human avatars rather than stylized avatars. Many computer vision and graphics techniques have been historically developed for real-world images, thus researchers naturally extend them to 3D realistic human avatars, benefiting from extensive available resources and benchmarks. Among methods developed specifically for 3D anime-style avatars, authors focus on single-view or multi-view image-based inputs, facing limitations in customization and flexibility compared to text-prompt-based approaches developed for 3D realistic human avatars. 

An interesting research path is to develop methods that are text-prompt-based, which could bridge the gap between 3D realistic human avatar generation, where text-prompt-based approaches have become dominant, and 3D anime-style avatar creation that currently relies heavily on image-based methods. Although text-prompt-based approaches have proven highly effective for 3D realistic human avatars, their application to anime-style remains underexplored. The shift from image-based to text-prompt-based approaches for 3D anime-style avatar generation, pivoted by AniDream~\cite{yamada2025anidream}, marks a significant change. This direction has the potential to build on the demonstrated effectiveness of text-to-3D models for realistic human avatars, while also addressing the distinct aesthetic requirements of anime. Text-prompt-based methods could also make 3D anime-style avatar generation more accessible and intuitive. Instead of needing a reference image of the anime character, users could describe abstract qualities like physical traits, clothing, or accessories. This also facilitates iterative refinement through progressive textual modifications, enabling adjustment of specific attributes without requiring new reference images or artistic skills. In the following, we discuss emerging trends and outline key open problems that must be addressed in the field.

\textbf{Fine-Tuning of Diffusion Models:} One of the most fundamental challenges in 3D avatar generation is obtaining high-quality, diverse datasets that adequately represent characters from various styles, perspectives, and contexts. Inconsistencies in training data propagate through to the generated results, causing issues with style coherence and character fidelity. While diffusion models are typically trained on large general datasets, fine-tuning these models for 3D anime-style avatar generation presents unique challenges. Current approaches using Low-rank Adaptation (LoRA) model show promising results but struggle with maintaining consistent character identity across novel poses and perspectives not seen in the fine-tuning data. Therefore, an interesting future direction is to explore advanced fine-tuning strategies, such as multi-stage fine-tuning, where the model is progressively exposed to more complex and diverse poses and expressions.

\textbf{Resolution and detail preservation:} Current generative models can encounter difficulties in preserving fine details and texture quality at higher resolutions, particularly for anime-style characters with distinctive line work or complex textures. This difficulty stems from a fundamental trade-off between global structure and local detail. Models must allocate their computational resources to maintain overall structural coherence in shape and proportions while also preserving fine-grained textures and resolution, often at the expense of one another. When models fail to preserve sufficient fine-grained structures, the resulting 3D anime-style avatars may lose sharp details or exhibit distorted facial features such as misaligned eyes or blurred expressions. These issues significantly reduce both recognizability and visual appeal. To address this limitation, researchers have employed multi-resolution processing strategies. In particular, one effective approach is hierarchical feature extraction, which processes the input at multiple scales simultaneously. Conceptually, this approach draws inspiration from the multi-scale representations used in mipmap pyramids within traditional graphics. Coarser levels capture the overall shape and structure, while finer levels preserve detailed textures and sharp boundaries. By combining information from these different levels, the model can better balance global coherence with local detail fidelity. This approach ensures that both the overall proportions of the character and the intricate visual details are accurately represented in the final 3D model.

\textbf{Limits of parametric models:}  
Models such as SMPL and SMPL-X are made for realistic human proportions, limiting their application to 3D anime-style avatars. 3D avatars that distance themselves from the parametric model present problems during the generation and rigging processes. As an example, methods that are developed for anime-style usually do not use parametric models, thus requiring custom rigging solutions that sacrifice transferability and standardization. Therefore, developing a more flexible parametric model that accommodates a wider range of artistic styles while maintaining rigging consistency and interoperability remains an open challenge.

\textbf{Avatars for animals or fantastic characters:} Non-humanoid creatures present unique challenges due to their diverse anatomy, movement patterns, and physical characteristics. This anatomical diversity is particularly significant given that the anime medium extensively features non-humanoid companion creatures and mascot characters. Current models struggle with the wide variety of animal forms, magical creatures, or fantasy beings that do not conform to human anatomy. The problem is intensified by limited training data for these subjects and the complexity of rigging non-standard skeletons for animation. Some methods dedicated to extracting the 3D models of different animals can be found in the literature, but they face the same issues as human-based parametric models regarding the representation of cartoon or anime-style characters. Therefore, the development of avatars for anime-style animals and fantasy characters remains an open problem. 

\textbf{Animation of hair and loose clothes:}  
In current frameworks for 3D anime-style avatar generation, physics-based animation often struggles to reproduce the characteristic motion of hair and clothing. Although specialized methods for animating anime-style hair and clothes exist, they are not integrated into current avatar generation frameworks. A further drawback is that incorporating such methods would significantly increase computational cost.

\textbf{Animation with detachable objects:}  
The exploration of interactions with objects, such as weapons or energy effects, is still in its early stages, demanding further investigation. However, the handling of combat powers like energy beams or projectile weapons presents unique challenges in 3D avatar generation and animation. These elements often require special treatment as they transition between being part of the 3D avatar model and independent objects in the scene.

\textbf{Benchmark for well-known anime characters:}  
There is a clear need for standardized evaluation datasets featuring diverse, well-known anime characters. Without such benchmarks, it remains difficult to objectively compare different generation approaches or measure progress in the field. Existing evaluations often rely on cherry-picked examples that may not represent the true capabilities or limitations of the frameworks. Therefore, proposing a benchmark, especially for text-prompt-based approaches, represents an ongoing challenge in the field.

\textbf{Evaluation protocol for user recognition:}  
Most user studies only ask about visual quality, not whether viewers can recognize the character. Future evaluations should include recognition tests to measure whether the 3D avatar keeps the key features of the target character. An automatic evaluation via pattern recognition should also be implemented into the assessment.

\textbf{Ethical and societal implications:}  
The use of generative 3D models, particularly for 3D anime-style avatar generation, presents several ethical challenges that need careful consideration. One of the primary concerns is content ownership. As generative models become more capable of producing highly detailed and lifelike 3D avatars, questions regarding intellectual property rights and ownership become increasingly complex. In the context of 3D anime-style avatars, this issue is further aggravated by the fact that the training datasets may include copyrighted material, such as artwork from well-known anime series, without explicit permission or compensation to the original creators. Addressing these concerns requires not only legal protections for creators and users but also a broader societal discussion about the ethical boundaries of generative technologies.

Looking forward, we consider that the field of 3D anime-style avatar generation is entering an important turning point. The continued reliance on image-based methods while 3D realistic human avatar generation surges ahead with text-prompt-based approaches is no longer justifiable. This reliance limits accessibility for creators without artistic skills and forces users into a rigid reference-matching workflow, thus constraining creative exploration that text prompts naturally afford. We believe the community must prioritize the development of text-prompt-based frameworks specifically designed for anime aesthetics, rather than treating 3D anime-style avatars as an afterthought in realistic avatar pipelines. Meanwhile, we must confront the ethical implications of training on copyrighted content. This concern weighs more heavily on 3D anime-style avatar generation, as anime characters are copyrighted intellectual property owned by specific creators and studios. Therefore, we advocate for the establishment of a more coordinated research direction that considers 3D anime-style avatar generation as a distinct task, rather than simply an adaptation of 3D realistic human avatar methods.

\section{Conclusion}

Over the past few years, notable advances in 3D anime-style avatar generation have been achieved. Despite this significant progress, several fundamental challenges persist in the field. One of the main difficulties involves preserving fine-grained details and textures at higher resolutions. Current frameworks often struggle with maintaining these details without increased computational costs. Another persistent challenge related to performance lies in simulating realistic hair and clothing behaviors. Physics-based animation approaches are computationally intensive, especially for real-time applications.

While challenges remain, the continued progress in 3D anime-style avatar generation is encouraging. Looking ahead, bridging the gap between 3D realistic human avatars and 3D anime-style avatars remains an important research direction. Although text-prompt-based approaches have achieved notable success with 3D realistic human avatars, their adaptation to the visual and structural characteristics of anime-style remains underexplored. Promising strategies include fine-tuning diffusion models with training data that captures stylized aesthetics, incorporating semantic controls to guide specific attributes such as hairstyle or clothing, and embedding anime-specific priors to enhance visual consistency. These directions can improve the fidelity and controllability of text-prompt-based 3D anime-style avatar generation, making the process more flexible and accessible for non-expert users.

\section{Acknowledgment}

This Project was conducted with the support of the Industrial Technology Innovation Program (20023347 Development of Graph-based Intelligent Metaverse Engine for Immersive Content-sharing Service) funded by the Ministry of Trade, Industry \& Energy of the Republic of Korea.

\def\refname{References}
\bibliographystyle{IEEEtran}
\bibliography{bib}

@inproceedings{yamada2025anidream,
  title     = {AniDream: Generating Skeleton-Guided Anime Avatars from Text Prompts},
  author    = {Fernanda Miyuki Yamada and Hiroki Takahashi},
  booktitle = {IEEE International Symposium on Mixed and Augmented Reality (ISMAR)},
  year      = {2025}
}

@article{huang2024dreamwaltzg,
  title={{DreamWaltz-G: Expressive 3D Gaussian Avatars from Skeleton-Guided 2D Diffusion}},
  author={Huang, Yukun and Wang, Jianan and Zeng, Ailing and Zha, Zheng-Jun and Zhang, Lei and Liu, Xihui},
  journal={arXiv preprint arXiv:2409.17145},
  year={2024},
  doi={10.48550/arXiv.2409.17145}
}

@article{loper2023smpl,
author = {Loper, Matthew and Mahmood, Naureen and Romero, Javier and Pons-Moll, Gerard and Black, Michael J.},
title = {{SMPL: A skinned multi-person linear model}},
year = {2015},
issue_date = {November 2015},
journal={Seminal Graphics Papers: Pushing the Boundaries},
publisher = {Association for Computing Machinery},
address = {New York, NY, USA},
volume = {34},
number = {6},
issn = {0730-0301},
doi = {10.1145/2816795.2818013},
month = oct,
articleno = {248},
numpages = {16}
}

@inproceedings{pavlakos2019expressive,
  title={{Expressive body capture: 3D hands, face, and body from a single image}},
  author={Pavlakos, Georgios and Choutas, Vasileios and Ghorbani, Nima and Bolkart, Timo and Osman, Ahmed AA and Tzionas, Dimitrios and Black, Michael J},
  booktitle={Proceedings of the IEEE/CVF conference on computer vision and pattern recognition},
  pages={10975-10985},
  year={2019}, 
  doi={10.1109/CVPR.2019.01123}
}

@article{huang2023dreamwaltz,
  title={DreamWaltz: Make a scene with complex 3D animatable avatars},
  author={Huang, Yukun and Wang, Jianan and Zeng, Ailing and Cao, He and Qi, Xianbiao and Shi, Yukai and Zha, Zheng-Jun and Zhang, Lei},
  journal={Advances in Neural Information Processing Systems},
  volume={36},
  year={2024},
  doi={10.5555/3666122.3666324}
}

@inproceedings{lin2022collaborative,
  title={Collaborative neural rendering using anime character sheets},
  author={Lin, Zuzeng and Huang, Ailin and Huang, Zhewei},
  booktitle={ International Joint Conference on Artificial Intelligence
AI and Arts},
  year={2023},
  pages = {5824-5832},
  doi={10.24963/ijcai.2023/646}
}

@inproceedings{cao2024dreamavatar,
  title={DreamAvatar: Text-and-shape guided 3D human avatar generation via diffusion models},
  author={Cao, Yukang and Cao, Yan-Pei and Han, Kai and Shan, Ying and Wong, Kwan-Yee K},
  booktitle={Proceedings of the IEEE/CVF Conference on Computer Vision and Pattern Recognition},
  pages={958-968},
  year={2024},
  doi={10.1109/CVPR52733.2024.00097}
}

@inproceedings{wang2024nova,
  title={{NOVA-3D: Non-overlapped Views for 3D Anime Character Reconstruction}},
  author={Wang, Hongsheng and Zhou, Xinrui and Lin, Feng},
  booktitle={Proceedings of the 6th ACM International Conference on Multimedia in Asia Workshops},
  pages={1-7},
  year={2024},
  doi={10.1145/3700410.3702127}
}

@article{he2024stdgen,
  title={{StdGEN: Semantic-Decomposed 3D Character Generation from Single Images}},
  author={He, Yuze and Zhou, Yanning and Zhao, Wang and Wu, Zhongkai and Xiao, Kaiwen and Yang, Wei and Liu, Yong-Jin and Han, Xiao},
  journal={arXiv preprint arXiv:2411.05738},
  year={2024},
  doi = {10.48550/arXiv.2411.05738}
}

@article{peng2024charactergen,
  title={{CharacterGen: Efficient 3D character generation from single images with multi-view pose canonicalization}},
  author={Peng, Hao-Yang and Zhang, Jia-Peng and Guo, Meng-Hao and Cao, Yan-Pei and Hu, Shi-Min},
  journal={ACM Transactions on Graphics (TOG)},
  volume={43},
  number={4},
  pages={1-13},
  year={2024},
  publisher={ACM New York, NY, USA},
  doi = {10.48550/arXiv.2402.17214}
}

\end{document}